\documentclass[12pt,a4paper]{article}
\usepackage{amsmath,amssymb}
\usepackage{graphicx}
\usepackage[sort]{cite}
\newcommand{\daps}{\delta_{aps}}
\newcommand{\es}{\mathbf{e}_S}
\newcommand{\fsl}{\mathbf{F}^{pert}_{SL}}

\begin{document}

\begin{center}

{\Large\bf Simple calculation of the Moon apsides motion}

\vskip10mm

{\large V.\ V.~Nesterenko}

\vskip7.5mm

{\small\it Bogoliubov Laboratory of Theoretical Physics,
Joint Institute for Nuclear Research,\\
Dubna, 141980, Russian Federation}

\end{center}

\vskip5mm

\noindent
\centerline{\bf Abstract}

\vskip2.5mm

\centerline{%
\parbox[t]{150mm}{%
A simple and clear method is proposed to calculate the averaged motion of the
apsis line in the Moon orbit. The obtained result is $3^{\circ}1'12''$ for
the starry period of the Moon revolution around the Earth or
$40^{\circ}22'48''$ per year. The modern observed value of the latter
quantity is $40^{\circ}41'$ per year. In ``Principia'' Newton derived
$1^{\circ}31'28''$ for the Moon month and $20^{\circ}12''$ per year. That is
approximately two times less than the observable values. Unlike the Newton
approach we use a simple and obvious averaging of the Sun disturbing force
for the starry period of the Moon revolution around the Earth. The
applicability of the obtained formulae to satellites of other planets and to
the planets themselves is grounded. Comparing Newton's calculation with our
method we reveal the reason, rather convincing, that brings Newton to
inadequate result.
\\[2.5mm]
}}

\textbf{Keywords:}~\parbox[t]{127mm}{The Moon apsides motion, disturbing force of the Sun,
the orbits closed to circles, stability of these orbits, the apsides motion in
circular-like orbits, Newton's ``Principia''.}

\vskip12mm

\section{Introduction}
\label{Sect:Intro}

Beginning with Newton times the motion of the Moon apsides is
considered to be the problem
that is inadmissible by simple mathematical tools making only use of
the basic geometrical peculiarities  in relative positions of the
Moon, the Earth, and the Sun
 (see, for example,  Chap.\ IX, \S\S ~197, 198 in Ref.\ \cite{Mult};
 and  Chap.\ I, \S~6, Chap.\ II,
 \S ~1 in Ref.\ \cite{Sub}). However it is difficult  to explain and
 understand this well know opinion if one
 takes into account the following facts. Practically without handling analytical
 methods Newton substantiated the universal inverse squares law for
 gravitational attraction by demonstrating  efficiency of this law,
 first of all, in explanation of deviations from the Kepler laws
 \cite{Prin, Prin-r}. Doing in this way Newton successfully solved a series
 of problems concerning, in particular, the perturbations of the Moon
motion due to the Sun, i.e., the Moon inequalities: the motion of the
Moon nodes, oscillation of the Moon orbit inclination; Newton also
explained the variation, eviction, and other inequalities in the Moon
orbit.

We propose a simple and clearly evident calculation of the average motion
of the moon apsides  yielding $3^\circ 3'12''$ for the starry period of motion in weakly
the Moon revolution around the Earth or $40^\circ 22'48''$ for the year.
The present-day observable value of this quantity is $40^\circ 41'$
per year. In ''Principia'' Newton derived respectively $1^\circ 31'28''$
and $20^\circ 12'$ that is approximately two times less than
observable values. Unlike Newton approach we use a simple and obvious
averaging of the sun disturbing force for the starry period of the Moon
revolution around the Earth. This results in centrifugal force,
additional to the Earth attraction and proportional to $r$.

The layout of the paper is the following. In Sec.\ 2 the averaged force
of the Sun disturbing the Moon motion relative to the Earth  is found.
In Sec.\ 3  we investigate the apsides motion in weakly perturbed
circular orbits. Here the stability of such orbits is considered also.
The Section 4 is devoted to numerical calculation of the Moon apsides
motion. In Sec.\ 5 we justify the applicability  of the obtained formulae
to the satellites of other planets. In Sec.\ 6 Newton's calculation is
compared with our approach. Here we reveal the reason  that brings
Newton to the inadequate result. In Sec.\ 7 (Conclusion) the obtained
results are formulated briefly.

\section{Averaged force of the Sun disturbing the Moon \\
motion relative to the Earth}
\label{Sc.2}

The disturbing force of the Sun will be found in the following simplifying
assumptions: {\it i})~The  Moon orbit is a circle with the centre in the
Earth  (see Fig. 1), that is applicable in view of a rather small
eccentricity of the Moon orbit $1/20 $; {\it ii}) It is assumed also
that the Moon orbit around the Earth and the Earth orbit around the
Sun lie in the same plane (in the  ecliptic plane). Here we  neglect the angle
of $5^{\circ }9'$ between the planes of these two orbits; {\it iii})~The
distance from the Earth  (the centre of the Moon orbit) to the Sun
$ES=D=150\times 10^6$ km. It is much larger  than the radius  of the
Moon orbit $D'=380\times 10^3 $ km. Therefore we can believe that at
any point  of the Moon orbit the attraction force  between the Sun and
the  Moon is parallel to the analogous force between the Sun and the
Earth, i.\ e.\ to the line  $SE$ connecting the Sun and the Earth (see
Fig.\ 1)\footnote{The same simplifying assumption was accepted
by Newton also. However he adopted it not in consideration of the
apsides motion but in calculation of the Moon nodes motion:``Both in
this calculation  and in the next ones  I believe that all the straight
lines drawing from the Moon to the Sun are parallel to the line connecting
the Sun with the Earth, for in some cases their inclination as much
decreases all the actions so in other cases it increases that;  as for
us we are looking for the averaged  motion of the nodes neglecting
such small points that only prevent calculation.''
See Ref.\ \cite{Prin} Book III, Preposition XXX.}.
In view of assumptions {\it i}) $-$ {\it iii}) the
attractive force of the Sun applied to the Moon at arbitrary point
$L$ (see Fig.\ 1) can be represented in the following form
\begin{equation}
\label{2e1}
\mathbf{F} _{SL}=\gamma \frac{S}{(D+D' \cos \theta)^2} \mathbf{e}_{S}
=\gamma \frac{S}{D^2}\frac{\mathbf{e}_{S}}{[1+(D'/D)\cos \theta]^2}.
\end{equation}
Here $\gamma $ is the gravitational constant; $S$ is the mass of the
Sun; $\mathbf{e}_{S}$ is a unit vector parallel to the line $ES$ and
directed  from$E$ to $S$; the quantities $D$ and $D'$ were defined before;
 $\theta $ is polar angle of the point $L$ (see Fig.\ 1). In our consideration
 the term `force' denotes the gravitational force divided by the mass
 of the body subjected to this force. Thus our force has the dimension
 of acceleration. Confining ourselves to the two terms  in expansion
 of \eqref{2e1} with respect to ratio $D'/D$ we get
\begin{equation}
\label{2e2}
\mathbf{F}_{SL}=\gamma \frac{S}{D^2}\mathbf{e}_S - 2\gamma
\frac{S}{D^3}D'\cos \theta  \ \mathbf{e}_S {.}
\end{equation}
 The first term in Eq.\ \eqref{2e2}  exactly equals the Sun attractive
 force acting upon the Earth
\begin{equation}
\label{2e3}
\mathbf{F}_{SE}=\gamma \frac{S}{D^2}\es{.}
\end{equation}
 This force should be omitted,\footnote{It is worth noting that
 this step is bounded with transfer of the coordinate system to the Earth
 orbiting the common inertia centre of the Earth and the Moon. See Ref.\
 \cite{Appel} Ch. XI,  \S$\,$  235, \S$\,$  236. The respective centrifugal
 acceleration is $\omega^2  D'L/(E+L)= (1/81)\omega^2D'$,
where $E$ is the mass of the Earth, $L$ is the mass of the Moon, $\omega=
2 \pi /T', T'$ is the starry period of the Moon revolution around the Erth.
The multiplier $1/81$ enables one to disregard this acceleration.} because
we are interested in in Sun force disturbing the Moon motion around the
Earth (see Ref.\ \cite{Prin}, Axioms or laws of motion, Corollary VI).
So the Sun disturbing force is
\begin{equation}
\label{2e4}
\mathbf{F}^{pert}_{SL}=-2\gamma \frac{S}{D^3}D' \cos{\theta}\ \es{.}
\end{equation}

\begin{figure}[t]
\centerline{\includegraphics[width=120mm,clip]{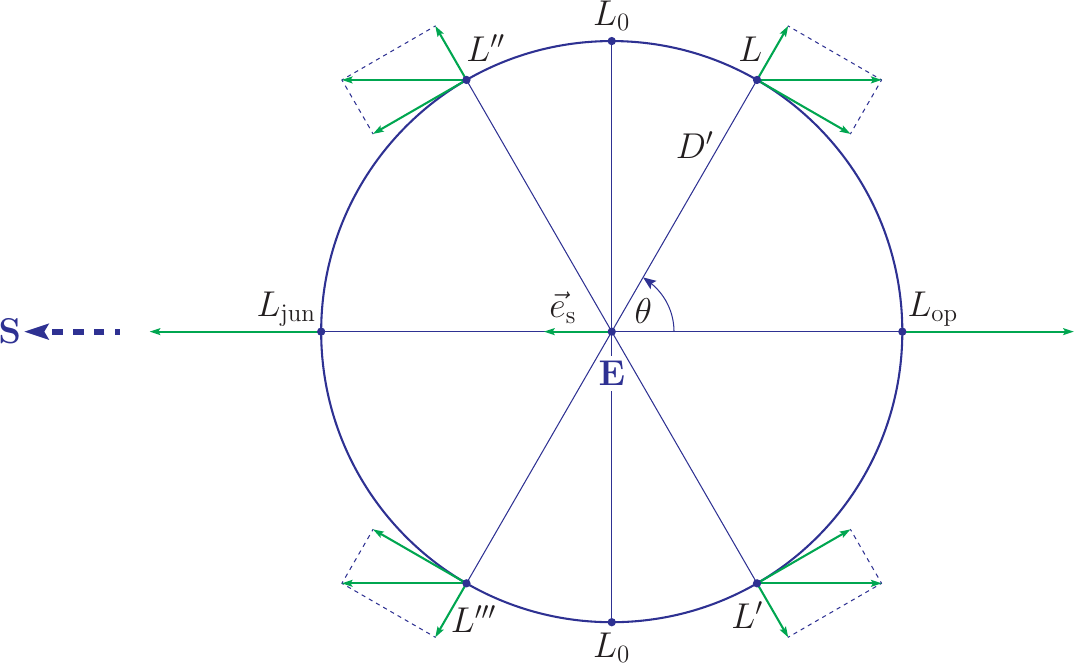}}
\caption{The Sun disturbing force, Eq.\   \eqref{2e4}, exerted on the  Moon
in circular orbit  (horizontal arrows). The other  arrows show the radial
and tangential  components  of the force.  Disturbing force  assumes
maximal absolute values at the points $L_{\rm{jun}}$ and
 $L_{\rm{op}}$  (at the moments  of junction and  opposition respectively).
The force directions  at these points are reverse.  The same takes
place  also at any two diametrically opposite  points  in the orbit (the
tidal character  of the disturbing force). At the points $L_0$ the disturbing
force vanishes. $E$ is the  Earth, $S$ is the Sun; $L, L'', L''', \ldots$
are different positions  of the Moon  in the orbit; $\es $ is  a unit vector
parallel  to the line $ES$ and directed from $E$ to $S$; $\theta $  is
the polar angle  of the arbitrary point $L$ reckoned  counter-clockwise
from the line $E L_{\rm{op}}$.}
\label{f1}
\end{figure}

Further we shall need the representation of the disturbing force \eqref{2e4}
in the polar coordinate system  $r,\theta$ with the pole at $E$ and
the polar axis $EL_{op}$. It can be obtained by making use of Fig.\ 1.
However it is much simpler to do in the following way. In the orthogonal
coordinate system with the origin at $E$, the $X$ axis directed along
the line $EL$, the $Y$ axis along the line  $EL_{op}$  the vector $\es$  has the
components $(-1,0)$.  Now we  employ the standard formulae defining the
transformation of the physical components of the vector from the
orthogonal coordinate system to the polar one (see, for example , Ref.\
\cite{KK} Chap.\ 6, \S$\, $6.5). It gives
\begin{equation}
\label{2e5}
-\es=-\cos{\theta} \, \mathbf{e}_r+\sin \theta \, \mathbf{e}_\theta{,}
\end{equation}
where  $\mathbf{e}_r$ and $\mathbf{e}_\theta $ are the unit  moving base
vectors in the polar coordinate system specified above. Thus in polar coordinates the
Sun disturbing force  \eqref{2e4} has the following representation
\begin{equation}
\label{2e6}
\fsl = 2\gamma \frac{S}{D^3} D'
(\cos^2{ \theta}\,\mathbf{e}_r - \cos{\theta}\sin{\theta}\, \mathbf{e}_\theta) {.}
\end{equation}

In Fig.\ 1 we can see that Eqs.\ \eqref{2e4}--\eqref{2e6} remain valid
in the IV quadrant also provided that the angle $\theta$ ranges from
$-\pi /2$ to  $\pi /2$ (i.\ e.\  the Moon moves from the point $L'$ to the
point $L$ in Fig.\ 1). Equations   \eqref{2e4}--\eqref{2e6} are true
also  when the Moon moves from the point $L''$ to the point $L'''$
lying in the II and in the III quadrants, respectively (the angle $\theta $
changes from $\pi/2$ to $3\pi/2$). Thus Eq.~\eqref{2e6}  defines
the polar coordinates of the Sun disturbing force for the whole period
of the Moon revolution around the Earth provided that the  angle $\theta$
ranges between $-\pi/2$ and $3\pi/2$.

In order to remove the dependence of the Sun disturbing force on the angle
$\theta$ we average  Eq.\ \eqref{2e6} over $\theta$ in the limits  $-\pi/2\le
\theta \le 3\pi /2$. For that we take advantage of the following integrals:
\begin{equation}
\langle \cos^2{\theta} \rangle  = \frac{1}{2\pi}
\int \limits^{3\pi/2}_{-\pi/2}\sin^2{\theta}\, d\theta =\frac{1}{4\pi}\int
\limits^{3\pi/2}_{-\pi/2}(\cos{2\theta} +1) \, d\theta =
\frac{1}{4\pi} \left(\left.\frac{\sin{2\theta}}{2}\right|^{\theta=3\pi/2}_{\theta=\pi/2}
+2\pi \right)=\frac{1}{2}{,}
\label{2e7}
\end{equation}
\begin{equation}
\langle \cos{\theta}\sin{\theta}\rangle  =\frac{1}{2\pi}\int\limits^{3\pi/2}_{-\pi/2} \frac{1}{2}
\sin{2\theta}\,d\theta=\left.\frac{1}{4\pi}\frac{\cos{2\theta}}{2} \right|^{\theta=\pi/2}_{\theta=3\pi/2}
=\frac{1}{8}\bigl(-1-(-1) \bigr)=0 \,{.}
\label{2e8}
\end{equation}
In these equations we can write  $d\theta =\omega dt$, where $\omega$
is the angular velocity of the Moon in circular orbit (for uniform motion
$\omega=2\pi/T, \, T$ is a period).  Thus the averaging \eqref{2e7} and
\eqref{2e8} over the Moon positions in the orbit  is equivalent to
the time-averaging. Asa result the averaged disturbing force of the
Sun, Eq.\ \eqref{2e6} acquires the form
\begin{equation}
\label{2e9}
\langle
\fsl
\rangle=\gamma \frac{S}{D^3}D'\mathbf{e}_ r\,{.}
\end{equation}

Finally in our approach the total central force determining the Moon
motion relative to the Earth is
\begin{equation}
\label{2e10}
\mathbf{F}_L(D')=\left (-\gamma \frac{E}{D'^2}+\gamma \frac{S}{D^3}D'  \right)
\mathbf{e}_r\,{.}
\end{equation}
The first term in the right-hand side of Eq.\ \eqref{2e10} is the
attraction force of the Earth exerted to the Moon, $E$ is the Earth
mass, and the second term is the disturbing force of the Sun is
represented as an additional, to the Earth attraction, centrifugal force
tending to move away (to separate) the Moon  from the Earth.

In deriving Eqs.\  \eqref{2e9} and \eqref{2e10} the circular orbit of
the Moon was envisaged. However it is obvious that these equations are
valid for all circular orbits but not only for the orbits  when $r=D'$.
In view of the smooth local dependence of the orbit shape on the the
acting force and, on the contrary, the same dependence  of the force
on the orbit  form on may expect that  Eqs.\  \eqref{2e9} and \eqref{2e10}
are also applicable, for example, to slightly disturbed  circular orbits,
to the elliptic orbits  with  small eccentricity and so on. The natural
measure of perturbation is, obviously, the ratio of the perturbation
force to the other forces acting on a body.  At the same  time this ratio
shows of the closeness of the disturbed orbits to the circular ones.
In the  problem at hand this measure is evidently the ratio of the
Sun disturbing force \eqref{2e9} to the Earth  attraction force (the
first term in Eq.\ \eqref{2e10} to the right).  With  $\kappa$ to denote
the absolute value  of this ratio we get
\begin{equation}
\label{2e11}
\kappa =\gamma \frac{S}{D^3}D':\gamma \frac{E}{D'^2}=\frac{S}{E}\left (
\frac{D'}{D}
\right )^3{.}
\end{equation}
This equation can be brought  into a more convenient form by making use
of the Kepler third law defined more exactly by Newton
\begin{equation}
\label{2e12}
\frac{S+E}{E+L}=\left(\frac{T'}{T}
\right)^2
\left(\frac{D}{D'}
\right)^3
\end{equation}
(see, for example, Ref.\ \cite{Appel} Ch.\  XI, \S\  236  or  Ref.\
\cite{Lamb} Chap.\ X, \S\ 81). Here the new notations are introduced,
namely: $L$ is the Moon mass (do not  mix with $L$ in Fig.\ 1);
 $T'=27.32$ days is the starry period of the Moon revolution around  the Earth;
 $T=365.26$ days in the starry period  of the Earth revolution around
 the Sun. In the left-hand site of of Eq.\  \eqref{2e12} we may neglect
 $L$ in comparison with $S$  and with $E$.  As a result  Eq.\ \eqref{2e12}
 becomes
\begin{equation}
 \label{2e13}
 \frac{S}{E}=
\left( \frac{T'}{T}
\right)^2
\left( \frac{D}{D'}\right)^3{.}
\end{equation}
 Substituting Eq.\ \eqref{2e13} into Eq.\ \eqref{2e11} we obtain
\begin{equation}
\label{2e14}
\kappa=\left(\frac{T'}{T} \right )^2 {.}
\end{equation}
The constant $\kappa $ can be represented also in the following form
\begin{equation}
\label{2ea14}
\kappa=\left(\frac{T'}{T} \right )^2=\left(\frac{T}{T'}\right)^{-2}=\frac{1}{N^2}{,}
\end{equation}
where $N=T/T'=13.37$ is the number of the moon revolutions around the Earth
during one year, i.\ e.\ the number of the lunar months  in year.

Numerically $\kappa$ is equal to
\begin{equation}
\label{2e15}
\kappa =\left(
\frac{27.32}{365.26}
\right)^2=\frac{1}{(13.37)^2}=\frac{1}{178.8}=5.582\times 10^3 {.}
\end{equation}
Approximately the same numbers are used in Ref.\ \cite{Lamb} Chap.\ X,
 \S\ 75; Chap.\ XI, \S\ 83.

\section{Apsides motion    in orbits close to circular ones}
\label{Sc.3}

 The idea of analytical calculation of the apsides motion and its
realization belongs rightfully to Newton (see Ref.\  \cite{Prin} Book I,
 Sec.\ IX, Proposition  XLV). In  the present-day setting it was
 employed in Ref.\  \cite{Lamb} Chap.\ XI, \S\S\ 87, 88 with the purpose
 to investigate stability  of weakly perturbed circular orbits.

 It is important to note that a circular orbit  is admissible  with
 {\it arbitrary law} of the attractive  central force $f(r)$ subject
 to  the condition that
 \begin{equation}
 \label{3e1}
\omega^2a=f(a){.}
 \end{equation}
Here $a$ is the radius of circular orbit, $\omega$ is the angular velocity,
$\omega=2\pi/T, T$ is revolution period.\footnote{The physical  meaning
 of Eq.\ \eqref{3e1} is simple.  The centrifugal acceleration of a body
 in circular orbit, i.e.\ the left-hand side  of Eq.\ \eqref{3e1},
 should be  equal to an external centripetal force, i.e.\ to the
 right-hand side of this equation.} From Eq.\ \eqref{3e1} it follows
 that the constant $h$ in the area law is to be determined by the  relation
 \begin{equation}
 \label{3e2}
 h^2=\omega^2 a^4=a^3  f(a){.}
 \end{equation}
 The general  equations of motion in polar coordinates $r, \theta$
 \begin{equation}
 \label{3e3}
 \frac{d^2r}{dt^2}-r \left(
 \frac{d\theta}{dt}
 \right)^2
 =-f(r),
 \quad r^2\frac{d\theta}{dt}=h,
 \end{equation}
 afterwards the  elimination of $(d\theta/dt)^2$
take on the form
\begin{equation}
\label{3e4}
\frac{d^2r}{dt^2}-\frac{h^2}{r^3}=-f(r) {,}
\end{equation}
Now we address  the orbits close  to the circular ones. It implies  that
 $r(t)$ in Eq.\ \eqref{3e4} can be represented in such a way
 \begin{equation}
 \label{3e5}
r(t)=a+x(t) {,}
 \end{equation}
 where   $r(t)$ is considered to be a small quantity. Let us substitute
 Eq.\ \eqref{3e5}  into Eq.\ \eqref{3e4}  and expand $f(r)$ into the
 Taylor series at the point $r=a$  confining ourselves  to the linear
 in $x(t)$ terms inclusive. As a result we get
\begin{equation}
\label{3e6}
\frac{d^2x}{dt^2} -\frac{h^2}{a^3}\left(1-\frac{3}{a}x
\right)
=-f(a)- f'(a) x {.}
\end{equation}
We can suppose that the relations \eqref{3e1} and \eqref{3e2} are
satisfied  as before though in the linear in $x(t)$ approximation. This
observation enables us to cancel the terms without $x(t)$ in Eq.\ \eqref{3e6}.
In consequence we arrive at {\it homogeneous} in $x(t)$ equation
\begin{equation}
\label{3e7}
\frac{d^2x}{dt^2}+\left[ f'(a) +\frac{3}{a} f(a)
\right] x(t)=0{.}
\end{equation}
It is exactly in this way the stability of non-linear equations is
explored (see, for example, Ref.\ \cite{KK} Chap.\ 9. Sec.\ 9.5 ){.}

  Solution $x(t)$ in Eq.\eqref{3e7} is bounded in magnitude, i.e.\ it
is stable, if the condition
\begin{equation}
\label{3e8}
n^2=f'(a) +\frac{3}{a}f(a) >0
\end{equation}
is fulfilled. Indeed, in this case the general solution of Eq.\ \eqref{3e7}
is a simple  harmonic oscillation
\begin{equation}
\label{3e9}
x(t)=C \cos{(nt+\varepsilon)}{,}
\end{equation}
where $C$ and $\varepsilon$ are the arbitrary real integration constants.
It is evident  that the absolute value of $x(t)$ does not exceed $|C|$.
So the general solution \eqref{3e9} is stable.

  Now we are ready to derive  directly formulae determining the apsides
  displacement in the orbit close to circular one  per one revolution of the body
   moving in this orbit.

In such orbits the  with the radius vector defined by  Eqs. \eqref{3e5}
and \eqref{3e9} the time required  in order that a body moves from the
to the point with  with minimal $r(t)$ to the point with  maximal $r(t)$
equals  $\pi /n$, where  $n$ is given by Eq.\ \eqref{3e8}. During this
period of time radius vector $r(t)$ turns  through the angle $\theta_{aps}$
\begin{equation}
\label{3e10}
\theta_{aps}=\frac{\pi}{n} \omega {.}
\end{equation}
It is clear that  $\theta_{aps}$   is an apsidal angle  in the orbit
slightly disturbed. Before substituting $n$ from Eq.\ \eqref{3e8} into
Eq.\ \eqref{3e10}  we a bit rearrange  Eq.\ \eqref{3e8}. We multiply
 respectively the left-hand side and the right-hand side of this equation
  by the equality following from Eq.\ \eqref{3e1}
$$
\frac{1}{\omega^2}=
\frac{a}{f(a)}{.}
$$
It gives
\begin{equation}
\label{3e11}
\frac{n^2}{\omega^2}=\frac{af'(a)}{f(a)}+3
\end{equation}
or in other form
\begin{equation}
\label{3e12}
\frac{\omega}{n}= \sqrt{\frac{f(a)}{3 f(a)+a f'(a)}}\, {.}
\end{equation}
Now we substitute Eq.\ \eqref{3e12} into  Eq.\ \eqref{3e10}. Finally
 the apsidal angle  $\theta_{aps}$ in the circular orbits slightly
 disturbed is determined by the expression
\begin{equation}
\label{3e13}
\theta_{aps}=\pi \sqrt{\frac{f(a)}{3f(a)+f'(a)}} \,{.}
\end{equation}

In order to account for theoretically inequalities in the Moon apsides
 line it is necessary to calculate the apsis displacement of the Kepler
 ellipse due to the Sun disturbance force. Let us find this displacement
 $\delta_{aps}$ for the Moon revolution for the Moon revolution
 around the Erth, i.e.\ during the Moon starry month. Keeping
 in mind  that the apsidal angle of the Kepler ellipse equals
 $\pi$ we can write
\begin{equation}
\label{3e14}
\daps=2(\theta_{aps}-\pi) {,}
\end{equation}
where $\theta_{aps}$ is apsidal angle in disturbed orbit. In our
approach, as well as in the Newton calculation (see Ref.\
\cite{Prin} Book I,
Sec.\ IX, Proposition XLV), the circular motion is substituted for
the motion in the Kepler ellipse and  $\theta_{aps}$is determined by
expression \eqref{3e13} which is the apsidal angle in the orbit close
to the circular one.\footnote{The observable disturbed orbit
of the Moon is like that $e=1/20$.}  As a result Eq.\ \eqref{3e14},
in virtue  of \eqref{3e13}, becomes
\begin{equation}
\label{3e15}
\daps=2\pi
\left(
\sqrt{\frac{f(a)}{3f(a)+f'(a)}}-1
\right){.}
\end{equation}
In what follows we shall use just this expression.

\section{Motion of the Moon apsides}
\label{Sc.4}

Now we have all the formulae necessary to numerical calculation of the
Moon apsides motion, namely, Eq.\ \eqref{3e15} determining the apsides
displacement for the Moon starry month and Eq.\ \eqref{2e10} for the
 resultant force exerted by the Earth and the Sun on the Moon. The latter
 expression  enables us to find the function $f(a)$  entering
 Eq.\  \eqref{3e15}. It is obvious that the radius $a$ in the circular
  orbit in Eq.\ \eqref{3e1} is the  radius of the Moon orbit $D'$. It
  is worthy to remind  that the function $f(a=D')$ introduced  in
  Eq.\ \eqref{3e1} is the centripetal force. That is why $f(D)=- F_L(D')$,
  where $F_{L}$ is the radial component of the total force in Eq.\
  \eqref{2e10}. Thus we get
  \begin{align}
  \label{4e1}
   f(D')&= \gamma \frac{E}{D'^2} -\gamma  \frac{S D'}{D^3}{,} \nonumber \\
   D' f'(D')&= -2 \gamma \frac{E}{D'^2} -\gamma  \frac{S D'}{D^3}{.}
  \end{align}
In virtue of this  the expression under the square root sign in
Eq.\ \eqref{3e15} can be represented in the form
\begin{equation}
\label{4e2}
 \left. \frac{f(a)}{3 f(a) + a f'(a)}\right|_{a=D'}=
\frac{\displaystyle \gamma \frac{E}{D'^2}-\gamma \frac{S D'}{D^3}}{\displaystyle \gamma \frac{E}{D'^2}-4 \gamma\frac{S D'}{D^3}}=
\frac{1-\kappa}{1-4 \kappa }{,}
\end{equation}
where $\kappa$ is the ratio
\begin{equation}
\label{4e3}
\kappa = \gamma \frac{S D'}{D^3} : \gamma \frac{ {E}}{{D'^2}} =
\left (\frac{T'}{T} \right )^2 {.}
\end{equation}
introduced  before (see Eqs.\ \eqref{2e11} and \eqref{2e14}). Finally
Eq.\ \eqref{3e15} defining the displacement of the Moon apsides for
one  revolution  around the Earth gives
\begin{equation}
\label{4e4}
 \daps = 2 \pi \left (
\sqrt{\frac{1-\kappa}{1-4 \kappa}} -1
 \right ){.}
\end{equation}
Evidently $\daps  >0 $  so the Moon  apsides motion is direct.

  The numerical value  of $\kappa$ obtained formerly  in Eq.\ \eqref{2e15},
$\kappa =1/(13.37)^2 \simeq 6\times 10^{-3}$,  allows us to limit
ourselves  in linear in $\kappa $ term in Eq.\ \eqref{4e4}
\begin{align}
\label{4e5}
\delta & = 2 \pi (1+3 \, \kappa +12\, \kappa ^2-16 \, \kappa ^3 \ldots )^{1/2}- 2 \pi  \nonumber \\
 &= 2 \pi  \left (
 \frac{3}{2} \kappa  +\frac{39}{8} \kappa ^2 \ldots
 \right )
= 3\pi \kappa \left (1 +\frac{13}{4} \kappa \ldots
\right ) \simeq 3 \pi \kappa {.}
\end{align}
In this equation and further the $\delta $ without  the superscript
denotes  the quantity $\daps$ obtained in the linear in $\kappa $
 approximation. In astronomy the degree measure of angles is used
 as a rule. In view of this we can write
\begin{equation}
\label{4e6}
\delta = 540^\circ \kappa {.}
\end{equation}
Substitution on the numerical value of $\kappa$, Eq.\ \eqref{2e15} , into
Eq.\ \eqref{4e6} gives
\begin{equation}
\label{4e7}
\delta = 540^{\circ} \kappa=540^{\circ}\frac{1}{(13.37)^2} =
3^{\circ}1'12''{,}
\end{equation}
that is approximately  two times  greater than the Newton result
 $1^{\circ}31'28''$ \cite{Prin} Book I, Sec. IX, Proposition XLV.

   Equation \eqref{4e7} permits us to derive  a simple expression for
 the motion of the Moon apsides line per year $\delta^{*}$
\begin{equation}
\label{4e8}
\delta^{*} = \delta \, N =  540^{\circ} \frac{1}{13.37}
= 40^{\circ}.38 = 40^\circ 22'48''{.}
\end{equation}
The contemporary value of this quantity obtained from observations is
$40^\circ 41'$. In view of the approximate character of our approach
 the obtained result, Eqs.\ \eqref{4e7}, \eqref{4e8}, should be recognized
 as a good one.\footnote{Our result  \eqref{4e8} is not only much more
  better, than the Newton's respective result, $20^{\circ}12'$ per year,
  but it is also better than the result $34^{\circ}22'$ due to  Clariaut
  (1749) obtained with allowance for the second order approximation.
  Our result is also better than the Newton result  $38^\circ 51'51''$
  discovered in his unpublished manuscripts (see Ref.\ \cite{Mult} Chap.\ IX, \S\,198
  and Ref.\  \cite{Sub} Chap.\ I, \S\,6; Chap.\ II, \S\,1).} The main
things  is that our result, with respect to accuracy, is completely
analogous to all other Newton's results concerning the the moon motion.

If in Eq.\ \eqref{4e5} we take into account the term proportional to
$\kappa^2$ then the following values are obtained: $\delta(\kappa ^2)=
3^{\circ}4'30'' $ (in place of Eq.\ \eqref{4e7}) and  $\delta^*(\kappa^2)=
41^\circ7'48''$ (instead of Eq.\ \eqref{4e8}). Without expansion into
 a series  in $\kappa$ formulae \eqref{4e4} yields $ \delta_{\rm{exact}}
 =3^{\circ}4'37''$ for  the Moon revolution around the Earth and
$ \delta^*_{\rm{exact}}=41^{\circ}8' 20'' $ per  year.
  Thus the approximation linear in $\kappa $ is quite acceptable
  in the problem in question.

  Closing this Section in is worthy to be
  convinced that stability of slightly disturbed circular orbits, Eq.\
  \eqref{3e8}, is satisfied in our calculation. Indeed after sub ostitution
 of the concrete form for the function $f(a)$ (see Eq.\ \eqref{4e1})
 into Eq.\ \eqref{3e8} the stability condition becomes
\begin{equation}
1-4 \kappa  > 0
\end{equation}
or
\begin{equation}
\label{4e9}
 \kappa  > \frac{1}{4} {.}
\end{equation}
In the case  of the Sun disturbing force  Eq.\   \eqref{2e15} this
restriction is wittingly  fulfilled.

\section{Applicability of the proposed approach to
the satellites of other planets and to the planets themselves}
\label{Sc.5}

In the first approximation the satellite orbits around the other
planets can be considered as slightly disturbed circular  orbits similar
to  the Moon orbit. Hence  Eq.\ \eqref{4e4}  is applicable  to calculating
the displacement  $\delta_{\rm{sat}}$ of apsides line  of any satellite
during its starry period $T_{\rm{sat}}$  of revolution around the planet
\begin{equation}
\label{5e1}
\delta_{\rm{sat}}= 2 \pi \left (
\sqrt{\frac{1-\kappa_{\rm{sat}}}{1-4 \kappa_{\rm{sat}}}} -1
 \right ){.}
\end{equation}
 In accordance  with Eq.\ \eqref{2e11} we can write
\begin{equation}
\label{5e2}
\kappa_{\rm{sat}}=\frac{S}{M_{\rm{pl}}}
\left(
\frac{a_{\rm{sat}}}{a_{\rm{pl}}}
\right )^3 {.}
\end{equation}
Here $S$  is the Sun mass as before;  $M_{\rm{pl}} $ is the planet mass;
$a_{\rm{sat}}$ is the average  distance of the satellite  from the planet;
$a_{\rm{pl}}$ is the average distance  of the planet from the Sun.
Again  it is  possible  to apply the Kepler third law~ \eqref{2e12}
\begin{equation}
\label{5e3}
\frac{S+M_{\rm{pl}}}{M_{\rm{pl}}+ m_{\rm{sat}}}=
\left (
	\frac{T_{\rm{sat}}}{T_{\rm{pl}}}
\right )^2
\left (
\frac{a_{\rm{pl}}}{a_{\rm{sat}}}
\right )^3
\end{equation}
($m_{\rm{sat}}$ is the satellite mass). Obviously in the Solar
planet  system we can  neglect the planet mass  $M_{\rm{pl}}$ in
comparison with the Sun mass S $((M_{\rm{pl}}/S)<10^{-3})$ and
the satellite mass $m_{\rm{sat}}$ in comparison  with the planet mass
$M_{\rm{pl}}$.  Taking into account  this we deduce from \eqref{5e3}
\begin{equation}
\label{5e4}
\frac{S}{M_{\rm{pl}}} =\left (
\frac{T_{\rm{sat}}}{T_{\rm{pl}}}
\right )^2
\left (
\frac{a_{\rm{pl}}}{a_{\rm{sat}}}
\right )^3{.}
\end{equation}
After substitution  of \eqref{5e4} into \eqref{5e2} we get
\begin{equation}
\label{5e5}
\kappa_{\rm{sat}} = \left (
\frac{T_{\rm{sat}}}{T_{\rm{pl}}}
\right )^2 {.}
\end{equation}
Thus Eqs.\ \eqref{5e1}, \eqref{5e2}, and \eqref{5e3} afford a complete
solution of the problem to calculate the motion of the apsides line
in the orbit of any satellite in the Solar planet system. \footnote{It is
worthy to note that the formulae derived here are inapplicable solely to
Uranus satellites. The point is that  the orbits of these satellites are
practically  at right angle to the Uranus orbit around the Sun. Therefore
the Sun force disturbing  the motion of the satellites relative  to the
Uranus should be found anew.}

In the present work the apsides motion  only due to the Sun disturbing
force  is considered. But it is worthy to note that the base well-grounded
formula \eqref{3e15} is applicable  to any  slightly disturbed circular orbit.
 The usage  of this formula  requires only  to determination  of the
 ratio  of the disturbing central  force to the planet attraction
 force  exerted on  the satellite, i.e.\ the ratio $\kappa_{\rm{sat}}$
 should be  found. After that  the working formulae  analogous  to
 Eq.\  \eqref{5e1} can be derived. For example, the disturbing force
 may be caused by deviation of the planet attractive law from the
 inverse squares law by virtue of the planet oblateness.

 It is clear that the general equation \eqref{5e1} may be employed
 for calculation of the  apsis motion  (in this case the perigee motion)
 in an orbit of any given planet due to the disturbing force  of other
  planet. To that end  only the ratio of disturbing force  to the Sun
   attraction force exerted on a given  planet  is to be found, i.e.\
    $\kappa_{\rm{pl}}$ should be calculated.

    \section{Comparison with Newton's  calculation}
\label{Sc.4}
Here we shall try to reveal the reason of the Newton  failure  in his
calculation of the apsis  motion in the Moon orbit. From the  very
beginning  it must be stressed that we did not succeed in recovering
Newton's pertinent  reasoning in detail.  Nevertheless the general
picture in this  problem is on general clear.

And so from Ref.\cite{Prin}Book I, Section IX, Proposition XLV it
unambiguously follows that Newton  represents the constant
component  of the Sun disturbing force as the centrifugal force,
additional to the  Earth attraction force, and proportional to $r$.
Denoting the ratio of these forces  through  $c$  Newton
derives the expression for the  apsidal angle  $\theta_N$ in the Moon
orbit
\begin{equation}
\label{6e1}
\theta_N =
\sqrt{
\frac{1-c}{1-4c}
}\times 180^\circ  {.}
\end{equation}
In order to verify  that  \eqref{6e1} is really apsidal angle  it is
sufficient to   compare  our Eqs.\  \eqref{3e12}, \eqref{3e13}, and
\eqref{4e4}. From the  physical point of view the  Newton ration $c$
and our ration $\kappa $ are {\it identical}.

Further in Ref.\ \cite{Prin}  Book III, Propositions III, XXV, and XXX Newton
 in fact finds  {\it the correct value} of $c$, namely
\begin{equation}
\label{6e2}
178\frac{29}{40} = \frac{1000}{178\, 725}
\end{equation}
(cp.\  with our Eq.\ \eqref{2e15}). However  in Ref.\  \cite{Prin}
Book I, Section IX, Proposition XLV Newton announces  without
explanations\footnote{Clairaut  (1743) believed that  the theory
of the Moon perigee   motion  has received the  most vague  development  in
the Newton studies (see Ref.\ \cite{Sub} Chap. II, \S\ 1).}
\begin{equation}
\label{6e3}
c = \frac{100}{35\,745}\, {.}
\end{equation}
The announced value \eqref{6e3} is connected with \eqref{6e2} by the
{\it exact} relation ship
\begin{equation}
\label{6e4}
c = \frac{100}{35\,745} = \frac{1000}{178\,7215} \times \frac{1}{2} \, {.}
\end{equation}

  One may suppose, with a rather great probability, that Newton was
urged  on insertion of multiplier  $1/2$ into Eq.\ \eqref{6e4} by virtue
of his analysis  of the high  tides theory (see Ref.\  \cite{Prin} Book III,
Propositions  XXXVI and XXXVII) that unfortunately  turned out
erroneous. This Newton error was found out and analysed by Laplace
(``Celestial Mechanics", Book XIII). Just in the Newton theory of tides
there appears the ratio $9:5$. Here Newton tries  to use the observable
heights  of the Sun and of the Moon tides  in determination of the ratio
of the Sun Force $(S)$ moving see to the same  moon force $(L)$. He
obtains
\begin{equation}
\label{6e5}
\frac{S}{L} = \frac{1}{4.4815} \,{.}
\end{equation}
But Laplace calculation gives
\begin{equation}
\label{6e6}
\frac{S}{L} = \frac{1}{2.3533} \,{.}
\end{equation}

   Further Newton utilises the ration $9:5$ in consideration of the
apsis motion in the orbits  of the Jupiter  and Saturn satellites
(see Ref.\ \cite{Prin}  Book III, Proposition XXII). At first Newton
 explains here how  to find the apsis motion (i.\ e.\ the motion of the
 orbit apices)  proceeding  from the motion of nodes. Newton could
 obtain  correct value  for apexes motion  if he followed this way. But
 Newton writes further ``However the apsides motion found out in
 such a way must  be decreased approximately  in ratio $5$  to $9$
 or in round $1$ to $2$ due the reason stating of which  is out  of
 the place here".

    It  should be  noted also that in the tides theory  Newton derived
wrong  ratio  of the Moon mass to the  Earth mass as $1:39.788$
for  $1:81.375$.  Further in Ref.\  \cite{Prin} Book III, Proposition
VIII  Newton  calculated  the masses  of the planets  Jupiter, Saturn,
  and Earth respectively as $\frac{1}{1\,067}$, $\frac{1}{3\,021}$, and
  $\frac{1}{169\,282}$ (the sun mass is $1$). All these mass values
  differ  from the contemporary  values  by the multiplier  about $1/2$.
 Probably is is  by virtue of  too high value  of the Sun horizontal
 parallax taken  at Newton's  times to be  $10''30'''$. The respective
 present-day  value is $8''\!\!.8$. From here the understated  value
 of the Sun distance  from the Earth  accepted at that times  follows
 \cite{Lamb} Chap.\  X, \S\ 75.

The inferences arising from the comparison  of the Newton calculation
with our approach  can be briefly  summarized as follows. The Newton
formula \eqref{6e1} is central in the problem in the problem in question.
This formula determining the apsidal angle  is completely correct. However
the numerical value  of the ration $c$ accepted by Newton, Eq.\ \eqref{6e4},
is absolutely  groundless. The additional multiplier $1/2$ in Eq.\ \eqref{6e4}
is unnecessary (superfluous). Very likely  that Newton  himself  had
no clear basis  to introduce this multiplier  because he stated  that
 corresponding reason is out of the place  in ``Principia".

 \section{Conclusion}
\label{Sc.7}
Calculation of the  Moon apsides  proposed by us is  in agreement
 with the observed data well. Applicability of the derived formulae
 to satellites of other planets  and to the planets themselves  is grounded.
 Our method distinguishes due its simplicity  and at the same time it
 is  easy-to-grasp.  The method is quite accessible  and intelligible
  even  for physicists  without experience in the celestial
mechanics.\footnote{It is interesting to note that  Newton himself
considered the Moon theory  to be difficalt very much and, they say,
that  he, being in despair, has told to his friend Halley, ``that he has
a headache from it and that it so often  deprives him a sleep,
that  he has decided do not think about it more", \cite{Mult} Chap.\ IX,
Historical essay and Bibliography.}

   In pertinent literature the opinion  becomes consolidated  according to which
a quite complete  result  concerning the Moon
apsides motion can be derived  only with allowance of {\it the second
order approximation}.\footnote{This point of view originates in the Clairaut
works (1743, 1749, 1752) \cite{Sub} Chap.\ II, \S\ 1.} Our method affords
quite acceptable  value in {\it the linear approximation} with respect to
the Sun disturbing  force, Eqs.\  \eqref{4e7}, \eqref{4e9}.

    Our approach completely  rehabilitates  the physical idea that
is leading one in Newton's studies and proceeds from the postulate:
the cause of motion is force.

Practically anew we have derived    the central in the problem
 at hand  expression  \eqref{6e1}  determining the apsidal
angle.\footnote{As far as we are aware the derivation of this formula by
present-day methods  is absent  in  the contemporary
literature.} Thus we corroborate  correctness  of the analytical part
of Newton's calculation. At the same time we quite convincingly  show
why  Newton's result  turned out inadequate.

\end{document}